\documentclass[twocolumn]{aastex63}
\usepackage[utf8]{inputenc}
\usepackage{amsmath}

\received{2020 July 6}
\accepted{2021 February 11}

\shorttitle{Activity Scaling Knee}
\shortauthors{Lehtinen et al.}

\begin{document}

\title{A Knee-Point in the Rotation--Activity Scaling of Late-type Stars with a Connection to Dynamo Transitions}

\author[0000-0002-1657-1903]{Jyri J. Lehtinen}
\affiliation{Max Planck Institute for Solar System Research, Justus-von-Liebig-Weg 3, D-37077 G\"ottingen, Germany}

\author[0000-0002-9614-2200]{Maarit J. K\"apyl\"a}
\affiliation{Department of Computer Science, Aalto University, PO Box 15400, FI-00076 Aalto, Finland}
\affiliation{Max Planck Institute for Solar System Research, Justus-von-Liebig-Weg 3, D-37077 G\"ottingen, Germany}	      
\affiliation{Nordita, KTH Royal Institute of Technology and Stockholm University, Roslagstullsbacken 23, SE-10691 Stockholm, Sweden}

\author[0000-0003-1445-9042]{Nigul Olspert}
\affiliation{Department of Computer Science, Aalto University, PO Box 15400, FI-00076 Aalto, Finland}

\author[0000-0001-6948-4259]{Federico Spada}
\affiliation{Max Planck Institute for Solar System Research, Justus-von-Liebig-Weg 3, D-37077 G\"ottingen, Germany}

\begin{abstract}
The magnetic activity of late-type stars is correlated with their rotation rates. Up to a certain limit, stars with smaller Rossby numbers, defined as the rotation period divided by the convective turnover time, have higher activity. A more detailed look at this rotation--activity relation reveals that, rather than being a simple power law relation, the activity scaling has a shallower slope for the low-Rossby stars than for the high-Rossby ones. We find that, for the chromospheric Ca {\small II} H\&K activity, this scaling relation is well modelled by a broken two-piece power law. Furthermore, the knee-point of the relation coincides with the axisymmetry to non-axisymmetry transition seen in both the spot activity and surface magnetic field configuration of active stars. We interpret this knee-point as a dynamo transition between dominating axi- and non-axisymmetric dynamo regimes with a different dependence on rotation and discuss this hypothesis in the light of current numerical dynamo models.
\end{abstract}

\keywords{Late-type stars (909), Stellar activity (1580), Stellar magnetic fields (1610), Stellar rotation (1629)}

\section{Introduction}

The magnetic activity of late-type stars is known to be correlated with their rotation rate,
faster rotation leading to increased levels of non-thermal emission in the upper stellar atmospheres from the chromosphere to the corona \citep[e.g.,][]{Noyes1984Activity,Vilhu1984MagneticActivity}. This scaling is understood to be a consequence of dynamo action in the turbulent outer convective envelopes of the stars, where the efficiency of the magnetic field generation is governed by the rotation and the non-uniformities related to it \citep[see, e.g.,][]{Ch10}, thus leading to different levels of magnetic heating.

The rotation-governed scaling holds for both the observed activity \citep{Noyes1984Activity,Gilliland1985RotationActivity,Basri1987BinaryActivity} and magnetic fields \citep{Saar2001MagneticFields,Auriere2015GiantMagneticFields,Folsom2018FieldEvolution,Kochukhov2020HiddenFields} for stars with sufficiently low Rossby numbers, ${\rm Ro} = P_{\rm rot}/\tau_{\rm c}$, where $P_{\rm rot}$ is the rotation period and $\tau_{\rm c}$ the convective turnover time in the stellar convection zone. For faster rotation, both the activity \citep{Vilhu1984MagneticActivity,Pizzolato2003CoronalRotationActivity,Douglas2014ClusterActivity,AstudilloDefru2017MDwarfActivity,Newton2017MDwarfActivity,Wright2018MDwarfActivity} and magnetic fields \citep{Reiners2009FluxSaturation,Vidotto2014RotationMagnetism} become decoupled from rotation\footnote{In observational literature, this is commonly referred to as the ``saturation regime'' of magnetic activity. In dynamo theory, however, this term is used to refer to the stage of stellar dynamos, where magnetic field, after an initial exponential growth, levels off (saturates) due to the interplay of various nonlinearities. All observed active stars in the main sequence or the giant branch, even those with slow rotation, are in this nonlinear saturation regime.}, although recent studies indicate a weak rotation dependence even in this regime \citep{Reiners2014RotationActivity,Shulyak2019MDwarfFields,Magaudda2020RotationActivity}. We will call here the high and low Rossby regimes of the rotation--activity relation as the ``rotation-dependent'' (RD) and ``rotation-independent'' (RI) regimes of activity or magnetic field scaling, respectively.

\begin{figure}
\plotone{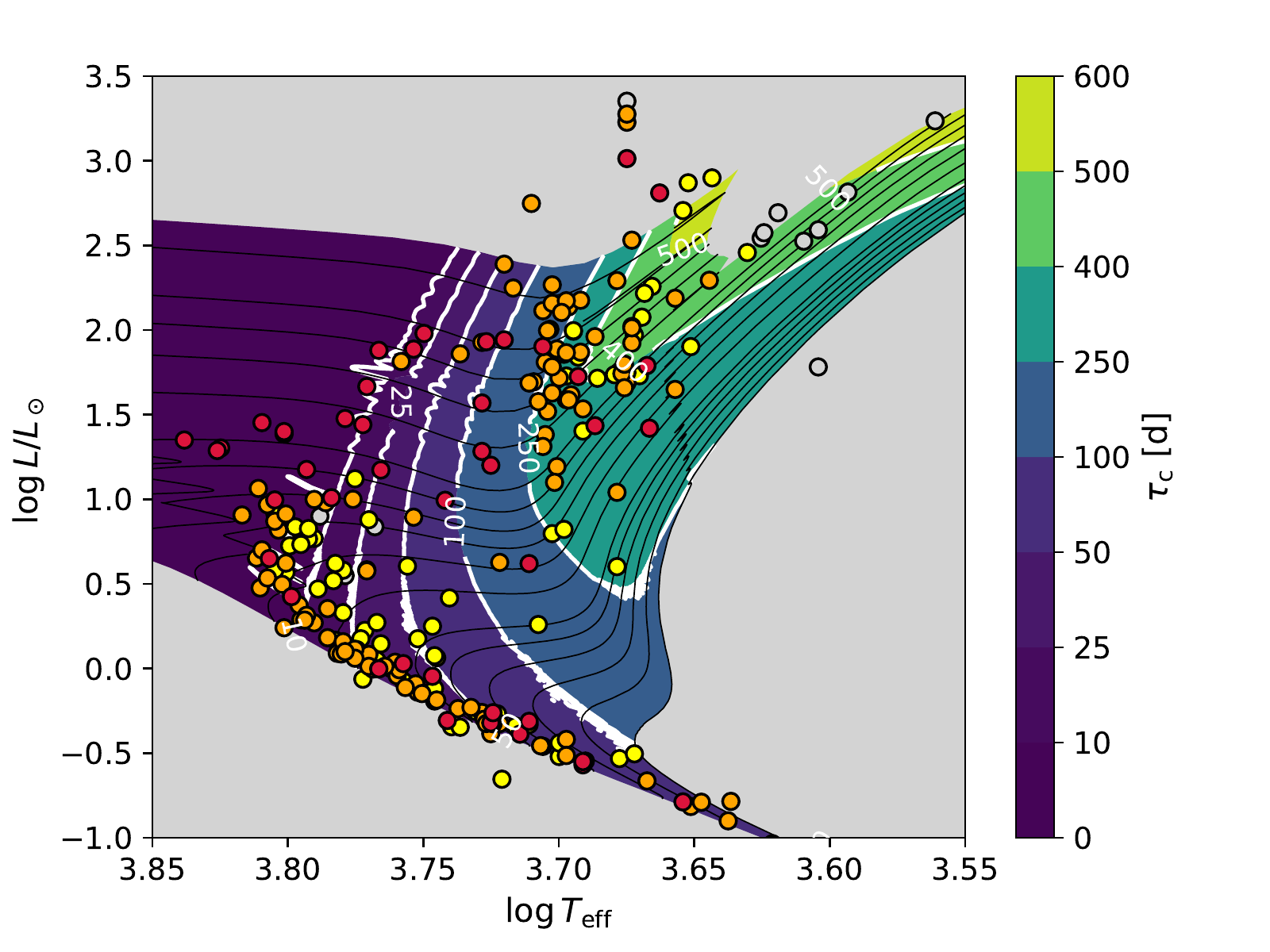}
\caption{Hertzsprung--Russell diagram of the stellar sample in \cite{Lehtinen2020RotationActivity}. Chromospheric activity of the stars is indicated by colour: \textit{red} for $\log R'_{\rm HK} > -4.4$, \textit{orange} for $-4.8 < \log R'_{\rm HK} < -4.4$, \textit{yellow} for $-5.2 < \log R'_{\rm HK} < -4.8$ and \textit{grey} for $\log R'_{\rm HK} < -5.2$. Evolutionary tracks and isocontours of the convective turnover time $\tau_{\rm c}$ are shown for solar metallicity.}
\label{fig:hr}
\end{figure}

Remarkably, both main sequence and evolved stars, at least in the RD regime, show indications of sharing a similar dynamo process. They follow the same rotation scaling of both activity \citep{Basri1987BinaryActivity,Lehtinen2020RotationActivity} and magnetic fields \citep{Auriere2015GiantMagneticFields,Kochukhov2020HiddenFields} and high activity stars are found both among the main sequence and evolved stars \citep[see][as well as Figure \ref{fig:hr} for the highest chromospheric Ca {\small II} H\&K fluxes, $\log R'_{\rm HK} > -4.4$]{Schroder2018GiantActivity}. Several slowly rotating giants also have distinct activity cycles \citep{Olspert2018Cycles}, providing clear evidence of the existence of cyclic dynamos in them.

Already \cite{Noyes1984Activity} and \cite{Rutten1987RotationActivity} noted that, also within the RD regime, the chromospheric rotation--activity relation is shallower for the faster rotators and has a break in its slope at mid-activity levels, around $\log R'_{\rm HK} \approx -4.5$. \cite{Noyes1984Activity} presented both an empirical polynomial and an exponential fit for the relation and speculated on the possible physical causes for its shape. They did not, however, come to a clear conclusion about the interpretation. Since then, the shape of the chromospheric activity scaling in the RD regime has most often been modelled by a smooth exponential \citep{Gilliland1985RotationActivity,Stepien1994RotationActivity,Kiraga2007RotationActivity,SuarezMascareno2016CycleRotation}, while some authors have also presented piecewise fits to account for a localised knee in the scaling relation \citep{Mamajek2008RotationActivity,Mittag2018RotationActivity,SuarezMascareno2016CycleRotation}. The same curved profile is also visible in the coronal X-ray scaling \citep{Vilhu1984MagneticActivity,Hempelmann1995CoronalRotationActivity,Mamajek2008RotationActivity,Mittag2018RotationActivity}, although this has not been studied in equal detail.

So far there has not been a conclusive interpretation of the shape of the activity scaling relation and none of the published studies present physically motivated justifications for their choice of fitting functions. Theoretically, power law relations, possibly with different exponents in different regimes, are the most naturally expected result from the magnetohydrodynamic (MHD) equations, wherefrom the non-dimensional numbers, like the Rossby number discussed in this study, are derived.\footnote{Theoretical studies usually prefer the Coriolis number over the Rossby number. These numbers are related to each other via inverse proportionality, $\rm Co \propto Ro^{-1}$.} A break between two power law segments with different slopes could relate to a transition from one dominating dynamo regime to another one with a different rotation dependence. No similar physical justification is available for the other proposed shapes of the scaling relation.

In this paper we follow on our previous study \citep{Lehtinen2020RotationActivity} of the chromospheric rotation--activity relation of partially convective main sequence and evolved stars with the aim of establishing a more physically motivated understanding of the observed knee. Our stellar sample covers the RD regime of the activity scaling up to the RD--RI transition, and is thus ideally suited for the study.

\section{Stellar data}

We use the same stellar data as in \cite{Lehtinen2020RotationActivity}, with a few additional constraints for the stellar selection. These data consist of the time averaged Ca {\small II} H\&K line core emission to bolometric flux ratios, $R'_{\rm HK} = F'_{\rm HK}/F_{\rm bol}$, and rotation periods, $P_{\rm rot}$, derived for the stars from the Mount Wilson Observatory $S$-index time series \citep{Wilson1978MWOHK}. he period search was performed using an initial search based on periodic Gaussian processes and fine tuning the period estimates and their errors using the Continuous Period Search method \citep{Lehtinen2012CPS}. We derived Rossby numbers for the stars using convective turnover times, $\tau_{\rm c}$, estimated from the YaPSI stellar evolution models \citep{Spada_ea2017}. The derivation of these stellar parameters is described in detail in \cite{Lehtinen2020RotationActivity} and the convective turnover times are discussed further in Sect. \ref{sect:tauc}. The data tables containing our rotation--activity data have been published by \citet{Lehtinen2020Data} and are described in Appendix \ref{sect:data}.

The subgiants have the most uncertain evolutionary track fits and thus cause considerable scatter in the rotation--activity relation. For this reason, we have excluded them from the current study. We have also excluded all stars with $\tau_{\rm c} < 5$~d, as was done in \cite{Lehtinen2020RotationActivity}, since these remain highly uncertain for all stars and are dominated by systematic errors stemming from the stellar structure models.

Finally, the original full sample included four stars (HD~3443, HD~3795, HD~22072 and HD~190360) with low activity but suspiciously short $P_{\rm rot}$ as estimated from the activity time series. We excluded all of these stars from our analysis as having spurious period detections. This interpretation was directly justified for HD~3443, HD~3795 and HD~22072, since their TESS or Kepler K2 photometry does not show any periodicity that could relate to their rotation. For HD~190360 there were no useful photometric time series available to check the validity of its $P_{\rm rot}$. However, we do not find any reason to think that it would in reality have such an anomalously short rotation period as suggested by the period analysis of the activity data.

Our final sample consists of 54 F- to K-type main sequence stars and 41 giants. We have analysed these separately for the main sequence sample (MS) and the full combined sample (Combined), which includes also the giants. This analysis allows for investigating the effect of the increased scatter from the giants in estimating the shape of the activity scaling relation. We have not analysed the giants as their own separate sample, since their larger scatter and concentration around the mid-activity levels, $-4.8 \lesssim \log R'_{\rm HK} \lesssim -4.4$, means that on their own they contain little information on the shape of the scaling relation.

\section{Results}

\subsection{Convective turnover times}
\label{sect:tauc}

\begin{figure}
\plotone{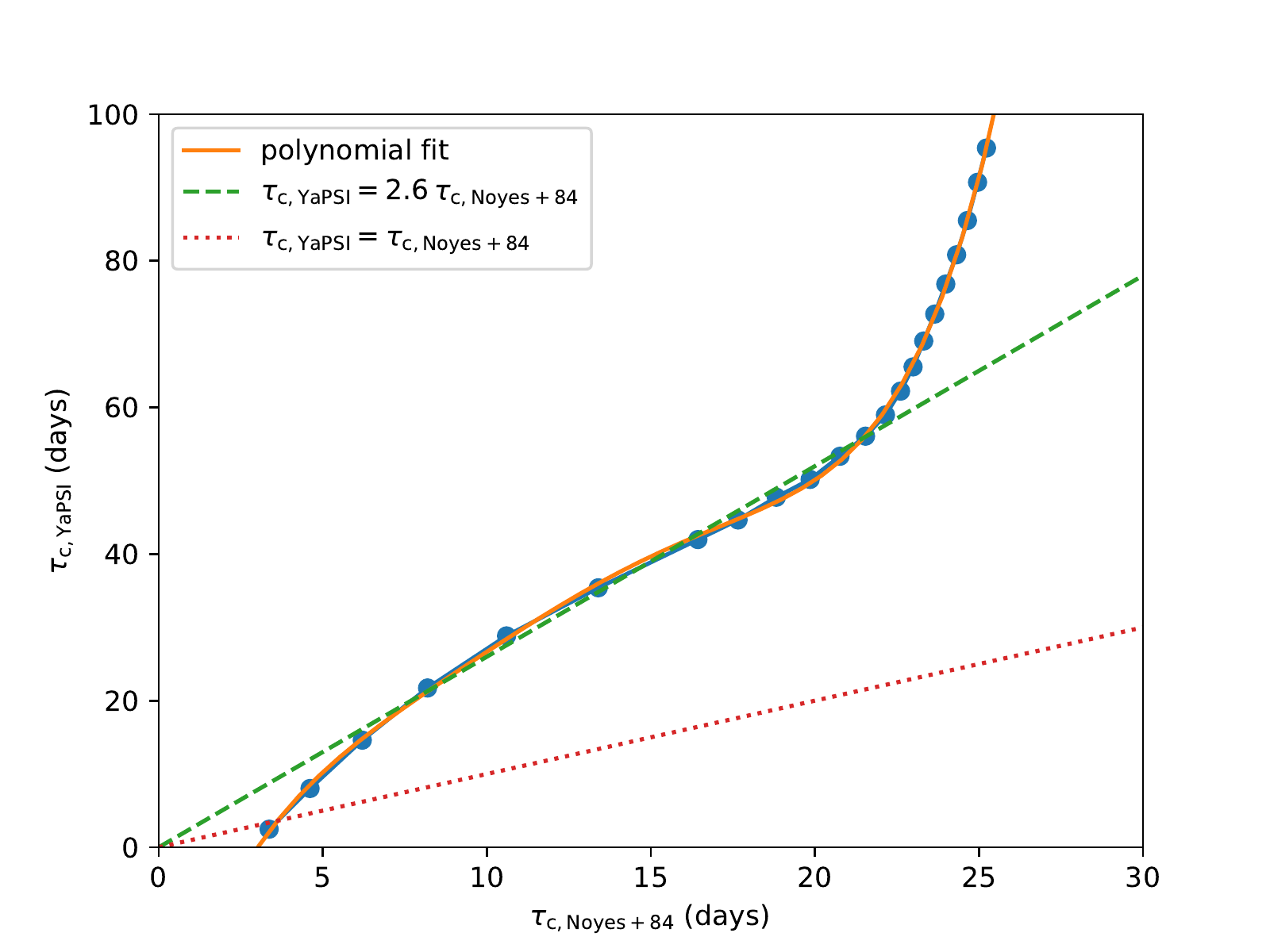}
\caption{Relation between the convective turnover times, $\tau_{\rm c}$, derived for main sequence stars from the YaPSI models and the \cite{Noyes1984Activity} empirical formula. The dotted and dashed lines indicate linear identity and $\tau_{\rm c,YaPSI} = 2.6\, \tau_{\rm c,N84}$ relations between the two $\tau_{\rm c}$ scales.}
\label{fig:tauc}
\end{figure}

The complex relation between the convective turnover time in the stellar outer convective envelope and the location of the star in the Hertzsprung--Russell diagram is illustrated in Figure \ref{fig:hr}. This shows, for solar metallicity, the isocontours of the turnover time $\tau_{\rm c,YaPSI}$, derived from the YaPSI stellar evolution models \citep{Spada_ea2017}. These convective turnover timescales were calculated from linearly interpolated evolutionary tracks at three different metallicities and then interpolated for the stellar metallicity. They are calculated as global averages over the convective envelope according to the global definition given in Appendix A of \citet{Spada_ea2013}.

The features of Figure \ref{fig:hr} reflect well-established results of stellar evolution theory.  Namely, low-mass and solar-like stars have convection zone throughout their evolution, which deepen significantly during the subgiant and red giant branch phase.  These stars therefore have convective turnover timescales of the order of a few days to about a hundred (depending on their mass) on the main sequence, increasing to several hundreds of days on the red giant branch. Intermediate-mass stars ($M \gtrsim 1.3\, M_\odot$) only acquire an outer convection zone after leaving the main sequence. During the subgiant phase, their convective turnover timescales increase rather abruptly from zero (when the outer convection zone is still absent) to values of the order of a few hundreds of days. These features of our model-based $\tau_c$ are in good qualitative and quantitative agreement with the calculations of \citet{Charbonnel_ea:2017}.

There is a steep increase in the turnover times towards the evolved giants, due to their greatly expanded outer layers. Such behaviour is not captured by the more commonly used ways of estimating $\tau_{\rm c}$, most notably the empirical formula by \cite{Noyes1984Activity}, which parameterizes $\tau_{\rm c}$ only for the main sequence as a function of the photometric color $B-V$. In general, for post-main sequence stars, $\tau_{\rm c}$ cannot be expressed as a function of one single parameter.

As the empirical formula of \cite{Noyes1984Activity} has been widely used, we have compared in Figure \ref{fig:tauc} their $\tau_{\rm c,N84}$ values with our $\tau_{\rm c,YaPSI}$. The $\tau_{\rm c,YaPSI}$ values were extracted from the stellar models at an evolutionary stage approximately halfway through the main sequence, defined as the instant when half of the central hydrogen has been exhausted by nuclear reactions. This choice, rather than a classical isochrone, is more representative of a heterogeneous sample of field stars, such as the ones included in the Mount Wilson catalog.

For a wide range of values ($\tau_{\rm c,YaPSI}=18$--$57$~d, and $\tau_{\rm c, N84}=7$--$22$~d, respectively, corresponding with late F- to early K-type main sequence stars), the turnover times follow approximately a linear relation, $\tau_{\rm c,YaPSI} = 2.6\, \tau_{\rm c,N84}$.
This approximation breaks down both at low and high $\tau_{\rm c}$, although the exact relation remains monotonous. For the higher mass main sequence stars, with the shortest $\tau_{\rm c}$, the $\tau_{\rm c,YaPSI}$ values quickly drop to zero as the outer convective envelope becomes shallower and disappears. For the low mass stars with the longest $\tau_{\rm c}$, the empirical $\tau_{\rm c,N84}$ fails to fully capture the steep increase, predicted by the models.

The relation between $\tau_{\rm c,YaPSI}$ and $\tau_{\rm c,N84}$ can be satisfactorily represented by a fifth-order polynomial fit,
\begin{equation}
\begin{split}
\tau_{\rm c,YaPSI} = & -30.7 + 15.2\tau_{\rm c,N84} \\
& - 2.16\tau_{\rm c,N84}^2 + 0.192\tau_{\rm c,N84}^3 \\
& - 0.00839\tau_{\rm c,N84}^4 + 0.000141\tau_{\rm c,N84}^5.
\end{split}
\label{eq:tauc}
\end{equation}
\noindent This relation, naturally, does not apply for the evolved stars, as $\tau_{\rm c,N84}$ is not defined for them.

\subsection{Two-piece power law model}
\label{sect:fit}

\begin{deluxetable*}{lccccc|ccc}
\tablecaption{Fit coefficients of the power law model for the MS and Combined samples\label{tab:fit}}
\tablehead{& $\log C_1$ & $\beta_1$ & $\beta_2$ & $\log {\rm Ro}_0$ & $\sigma$ & $\log R'_{\rm HK,0}$ & ${\rm Ro}_0$ & ${\rm Ro}_{0,\rm N84}$}
\startdata
MS & $-4.745^{+0.051}_{-0.047}$ & $-0.458^{+0.062}_{-0.061}$ & $-1.222^{+0.146}_{-0.181}$ & $-0.456^{+0.054}_{-0.068}$ & $0.084^{+0.009}_{-0.008}$ & $-4.536^{+0.051}_{-0.045}$ & $0.350^{+0.047}_{-0.051}$ & 0.91 \\
Combined & $-4.807^{+0.038}_{-0.036}$ & $-0.512^{+0.059}_{-0.059}$ & $-1.507^{+0.279}_{-0.293}$ & $-0.310^{+0.030}_{-0.050}$ & $0.119^{+0.010}_{-0.008}$ & $-4.647^{+0.039}_{-0.033}$ & $0.490^{+0.036}_{-0.053}$ & 1.3 \\
\enddata
\end{deluxetable*}

\begin{figure*}
\plottwo{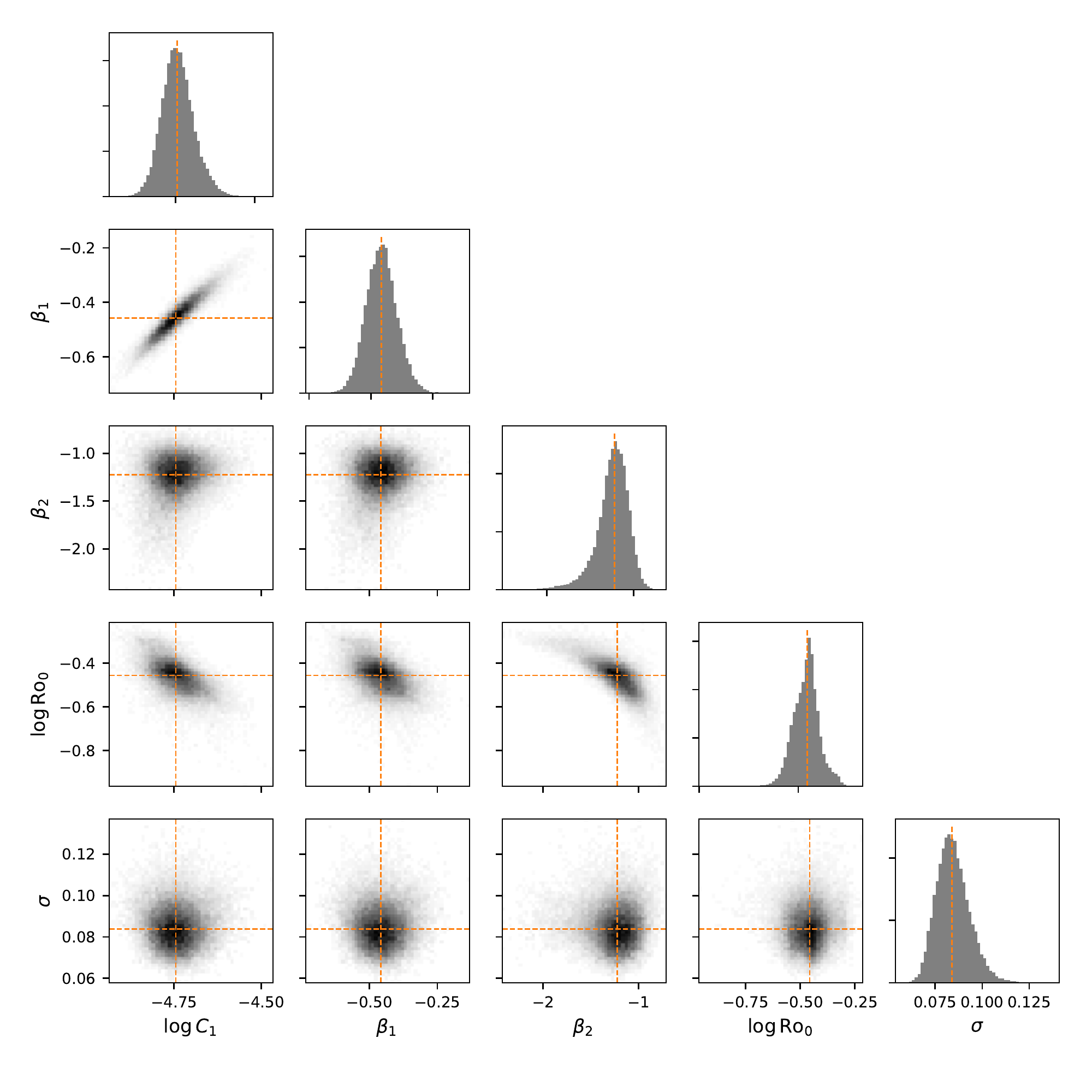}{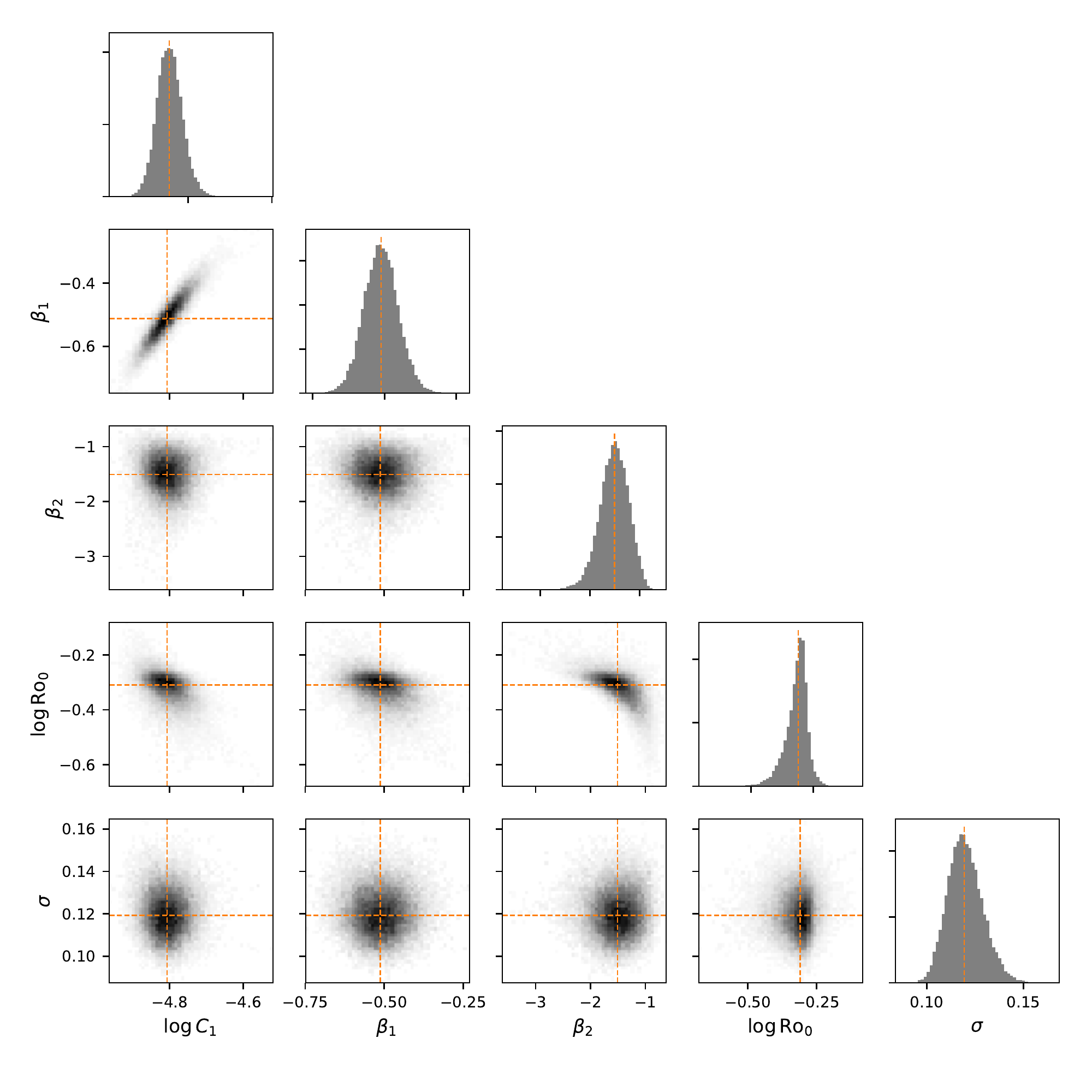}
\caption{The posterior distributions of the two-piece power law model parameters for the MS sample (left panel) and the Combined sample (right panel). The median estimates of each parameter are marked with the orange dashed lines.}
\label{fig:corner}
\end{figure*}

We modelled the shape of the observed $R'_{\rm HK}$ vs.~$\rm Ro$ rotation--activity relation using a two-piece power law model. This offers the most physically motivated description of the activity scaling, as discussed in the Introduction, and allows locating a precise knee-point for the scaling law, that can be compared with other results.

Our regression model is
\begin{equation}
R'_{\rm HK}({\rm Ro}) = f({\rm Ro}) =
\begin{cases}
C_1 \, {\rm Ro}^{\beta_1}, & {\rm Ro} \leq {\rm Ro}_0 \\
C_2 \, {\rm Ro}^{\beta_2}, & {\rm Ro} > {\rm Ro}_0,
\end{cases}
\label{eq:model}
\end{equation}
\noindent where $C_2 = C_1 \, {\rm Ro}_0^{\beta_1-\beta_2}$, ensuring continuity at the knee-point, $\rm Ro = Ro_0$. We assumed that the observed uncertainties of $\log R'_{\rm HK}$ follow a normal distribution, leading to a log-normal likelihood function for $R'_{\rm HK}$,
\begin{equation}
p(R'_{\rm HK}|\mu,\sigma^2) = \textrm{log-Normal}(R'_{\rm HK}|f({\rm Ro}),\sigma^2),
\label{eq:likelihood}
\end{equation}
\noindent where the mean, $\mu$, is given by the regression model $f({\rm Ro})$ and the scale by the scatter parameter $\sigma$. This analysis is similar to that of \cite{Douglas2014ClusterActivity}, \cite{Newton2017MDwarfActivity}, \cite{Wright2018MDwarfActivity}, and \cite{Magaudda2020RotationActivity} with the exception that our model does not cover the RI regime, but has instead separate power law exponents $\beta_{\{1,2\}}$ for the upper and lower parts of the RD regime, separated by $\rm Ro_0$.

We performed the model regression by treating $\log C_1$, $\beta_1$, $\beta_2$, $\log \rm Ro_0$, and $\sigma$ as the independent free parameters and sampling their joint posterior distribution using the \texttt{emcee} Python package \citep{ForemanMackey2013emcee}, which implements an affine invariant Markov Chain Monte Carlo (MCMC) ensemble sampler \citep{Goodman2010EnsembleSamplers}. We used weakly informative Gaussian priors $p(\beta) = \mathcal{N}(\beta|{-1},1^2)$ and $p({\log \rm Ro_0}) = \mathcal{N}({\log \rm Ro_0}|{-0.5},0.25^2)$ for $\beta_{\{1,2\}}$ and $\log \rm Ro_0$, a non-informative Jeffreys prior $p(\sigma^2) = 1/\sigma^2$ for $\sigma$ and a uniform prior for $\log C_1$. The $\beta_{\{1,2\}}$ and $\log \rm Ro_0$ priors were chosen to specify negative slopes for the fit and a rough knee-point location, based on visual inspection of the data. In addition, we discarded stars with $\rm Ro > 1$ from the fit, since this range is dominated by scattered outliers \citep[see further discussion in][]{Lehtinen2020RotationActivity}.

Since the internal uncertainties associated with the $R'_{\rm HK}$ and $\rm Ro$ values are much smaller than the scatter of the stars in the rotation--activity diagram \citep[see][]{Lehtinen2020RotationActivity}, we did not include these in our likelihood function Eq.~\ref{eq:likelihood}. Instead, the scatter parameter $\sigma$ only models the overall spread of the stars around the regression fit. Running the model regression with a modified likelihood function, which includes the $R'_{\rm HK}$ uncertainties, produced identical results with our main model to within error limits, thus demonstrating that the internal uncertainties have an insignificant effect to our results. This modified model regression is described further in Appendix \ref{sect:errfit}.

We set up the MCMC sampler to run with 100 chains of 2000 iterations and removed the first half of each chain to ensure good convergence. The parameter and error estimates were calculated as the medians and the 16th and 84th percentiles of the sampled Markov chains. The full posterior distributions for the MS and Combined samples are shown in Figure \ref{fig:corner}, including histograms for the marginal distributions of each model parameter.

\begin{figure*}
\plottwo{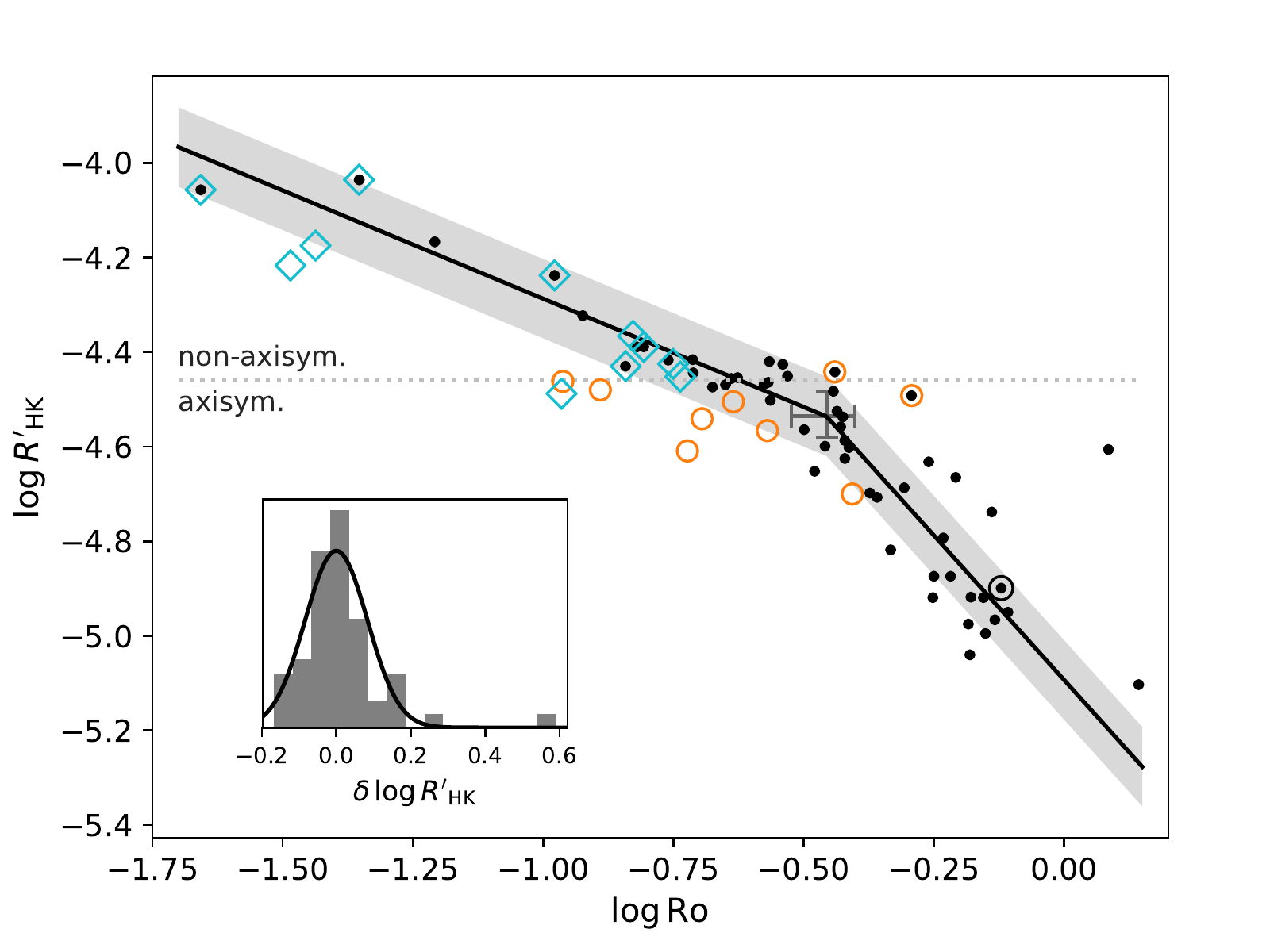}{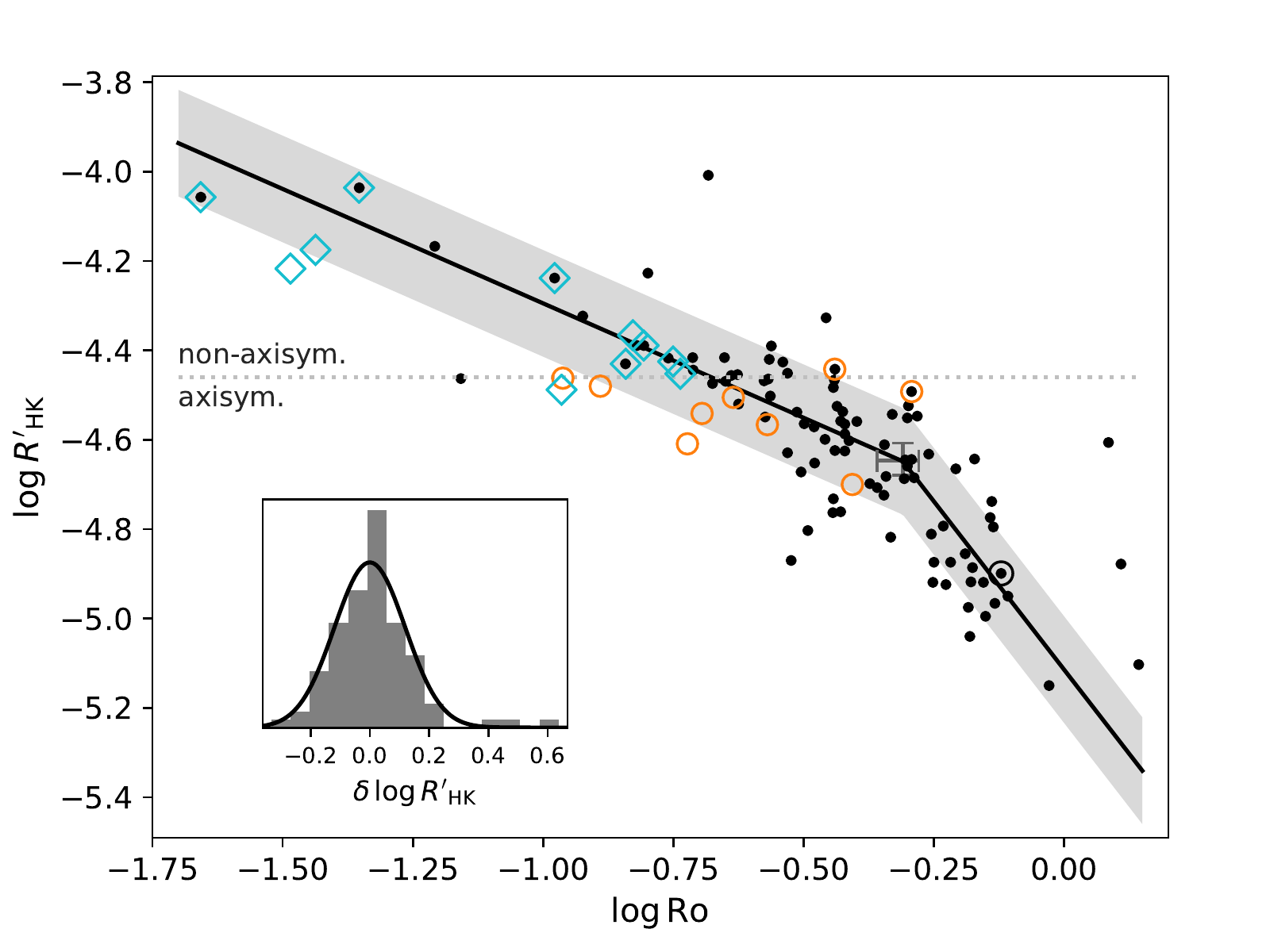}
\caption{Two-piece power law fits to the $R'_{\rm HK}$ vs.~$\rm Ro$ rotation--activity relation of the MS (left panel) and Combined (right panel) samples. The light grey shaded area marks the $\pm 1 \sigma$ range of the likelihood function and the error bars at the knee-point indicate the uncertainty in its location. The orange circles and cyan diamonds denote stars that \cite{Lehtinen2016Photometry} found to have axisymmetric or non-axisymmetric spot distributions, respectively, and the horizontal dotted lines mark the identified transition between these regimes. Insets show the $\delta \log R'_{\rm HK}$ residual distributions and the corresponding likelihood function profiles. The Sun is indicated as the black circled dot.}
\label{fig:powlaw}
\end{figure*}

The resulting parameter estimates are listed in Table \ref{tab:fit}, including derived estimates for the knee-point location as $\rm Ro_0$ and $\log R'_{\rm HK,0} = \log f({\rm Ro_0})$. We have also calculated rough estimates for the knee-point Rossby number in the \cite{Noyes1984Activity} scale, ${\rm Ro}_{0,\rm N84}$, to aid comparison with other published studies. These values were calculated from our ${\rm Ro}_0$ values in the YaPSI scale, using the approximate linear relation $\tau_{\rm c,YaPSI} = 2.6\, \tau_{\rm c,N84}$. A more accurate rescaling is not feasible, since the nonlinear relation between the two $\tau_{\rm c}$ scales (Eq.~\ref{eq:tauc}) does not directly translate for the Rossby numbers, which also depend on $P_{\rm rot}$. For the same reason we also do not attempt to derive accurate error estimates for ${\rm Ro}_{0,\rm N84}$.

In Figure \ref{fig:powlaw} we show the regression fits for the MS and Combined fits. They show general agreement with each other, although the Combined sample has increased scatter from the giants around the knee-point, which has pushed its $\rm Ro_0$ towards larger values. The larger $\rm Ro_0$ value leaves a narrower $\rm Ro$ range for the lower part of the two-piece model, which has caused the uncertainty of the $\beta_2$ parameter to increase in the Combined sample in relation to the MS sample. We find, nevertheless, that the better defined MS sample places the knee close to the $\log R'_{\rm HK} = -4.46$ limit where \cite{Lehtinen2016Photometry} found a sharp transition between axisymmetric and non-axisymmetric spot activity, surfacing as long-lived active longitudes. This suggests that the onset of non-axisymmetry and the break in the activity scaling slope may be related phenomena.

Further evidence supporting this claim is provided by a comparison with \cite{See2016FieldTopology}. They studied the magnetic topologies of a sizable sample of active stars from Zeeman Doppler imaging inversions and found a transition from mostly poloidal axisymmetric fields at slow rotation to mostly toroidal non-axisymmetric fields at fast rotation, occurring around their $\rm Ro = 1$. Their Rossby numbers were based on a $\tau_{\rm c}$ scale closely related to \cite{Noyes1984Activity}, so their transition line can be compared with our ${\rm Ro}_{0,\rm N84}$. Their axisymmetric to non-axisymmetric transition falls thus close to the knee-point in both our MS and Combined samples.

Insets in Figure \ref{fig:powlaw} show the residuals of $\log R'_{\rm HK}$ against the regression model,
\begin{equation}
\delta \log R'_{\rm HK} = \log R'_{\rm HK} - \log f({\rm Ro}).
\end{equation}
\noindent These are in both cases in good agreement with the profiles of the log-normal likelihood function (Eq.~\ref{eq:likelihood}).

Finally, we tested the validity of the two-piece power law model against the often used exponential model,
\begin{equation}
R'_{\rm HK}({\rm Ro}) = Ce^{\beta \rm Ro},
\label{eq:exp}
\end{equation}
\noindent using the same likelihood function (Eq.~\ref{eq:likelihood}). The parameter estimates for this exponential model are $\log C = -4.177^{+0.027}_{-0.027}$, $\beta = -1.065^{+0.062}_{-0.061}$ and $\sigma = 0.091^{+0.010}_{-0.008}$ for the MS sample and $\log C = -4.214^{+0.028}_{-0.029}$, $\beta = -0.968^{+0.067}_{-0.065}$ and $\sigma = 0.121^{+0.009}_{-0.008}$ for the Combined sample.

To compare the two models, we calculated the values of their relative Bayesian information criterion \citep[BIC,][]{Stoica2004ModelSelection},
\begin{equation}
\Delta {\rm BIC} = {\rm BIC}(\textrm{power law}) - {\rm BIC}(\textrm{exponential}).
\end{equation}
\noindent For the MS sample we found $\Delta {\rm BIC} = -2.89$, meaning that the two-piece power law model minimizes the BIC and provides a better model for the data. For the Combined sample we find the opposite to be true, $\Delta {\rm BIC} = 3.52$, which would favour the exponential model instead. This may be attributed to the larger scatter of the Combined sample, which makes the knee-point less defined. We claim here that the power law behavior is more physical and that the preference for an exponential model in the Combined sample results from the higher uncertainty in determining $\tau_{\rm c}$ for the evolved stars \citep{Lehtinen2020RotationActivity}.

\subsection{Gaussian clustering model}

\begin{figure*}
\plottwo{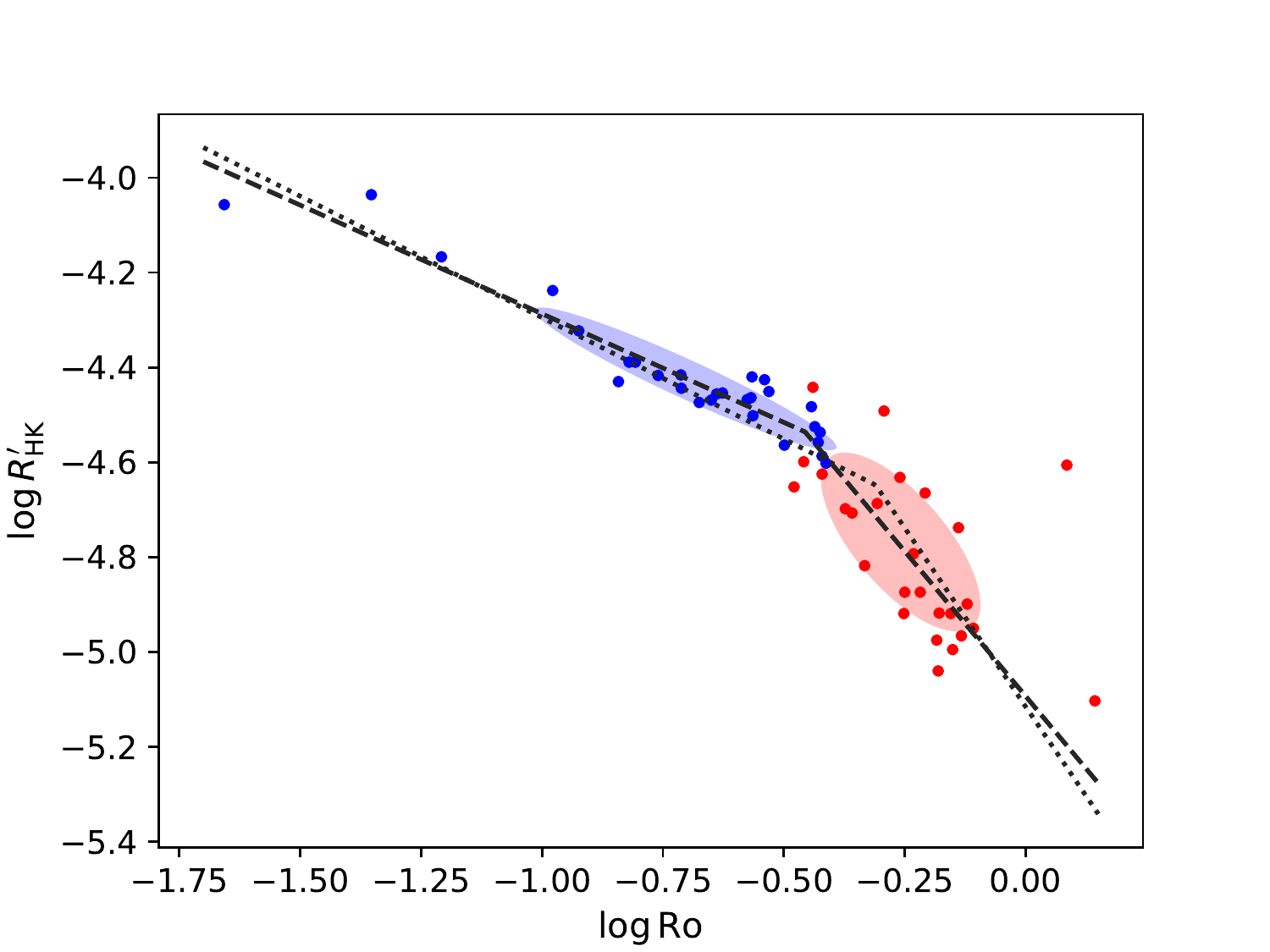}{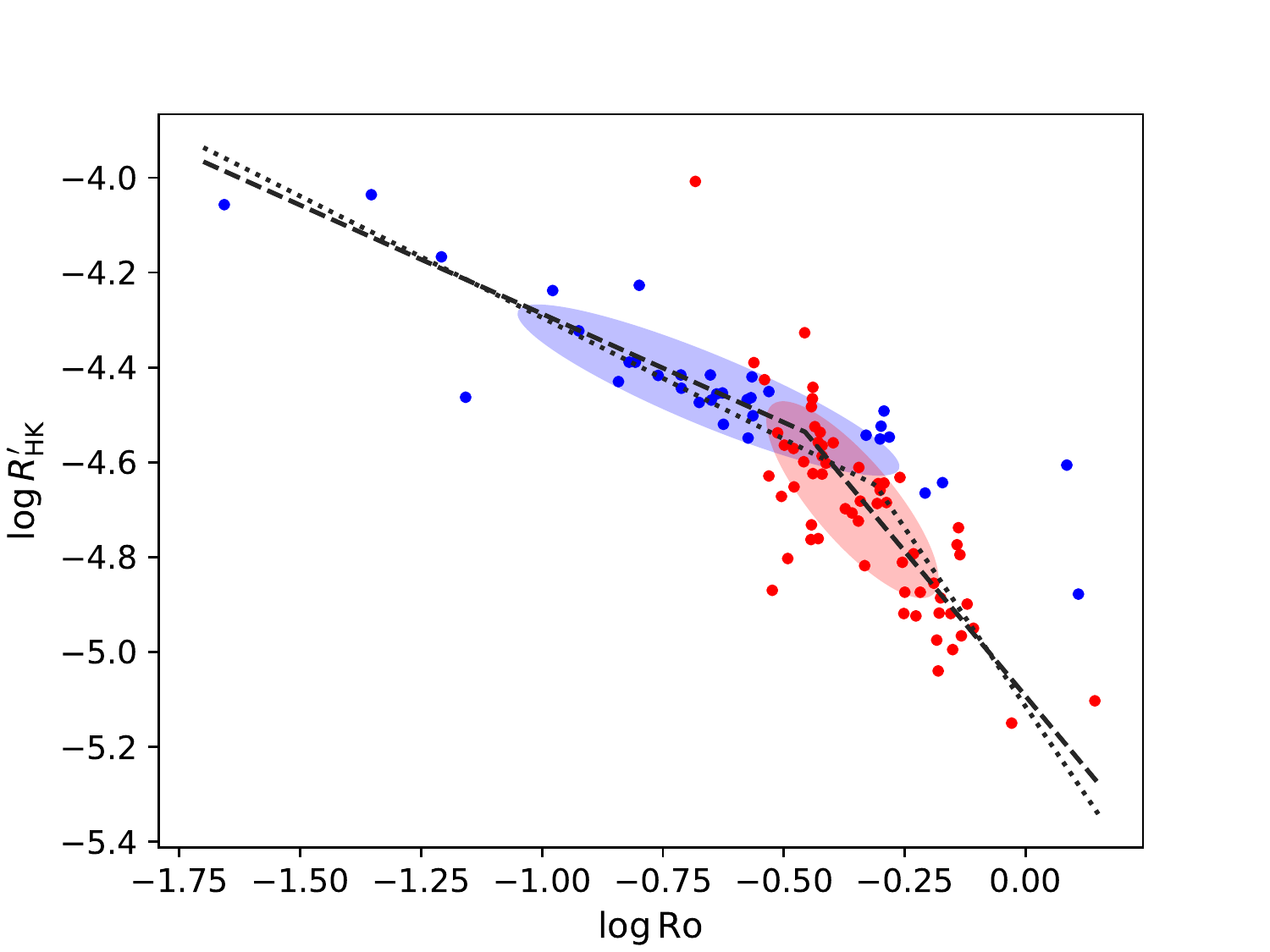}
\caption{Optimal Gaussian clustering for the MS (left panel) and Combined (right panel) samples. The clusters are denoted by the blue and red ellipses and their inferred members with corresponding colors. The regression fits of the MS (dashed line) and Combined samples (dotted line) are shown on top of the clusters.}
\label{fig:cluster}
\end{figure*}

To get a statistically independent look at the data, we also applied a Gaussian mixture model with expectation maximisation algorithm \citep{BarberBRML2012} for both the MS and Combined samples. We tested models with a number of clusters from one to five and determined the best model by minimizing their BIC.

The clustering results are shown in Figure \ref{fig:cluster}. For both the MS and Combined samples the data are best described by a bimodal model with two clusters intersecting at the the knee-point. This hints at an abrupt change in the scaling relation, inconsistent with a smooth exponential model. Note that in \cite{Lehtinen2020RotationActivity} we found a single Gaussian cluster for the whole RD regime and a surrounding outlier cluster, using the same algorithm. In the current study we have excluded the obvious outliers from the sample, explaining the improved ability of the clustering to model the shape of the activity scaling.

Figure \ref{fig:cluster} also includes the regression fits for the two samples. Notably, both for the MS and Combined samples, the two clusters intersect close to the knee in the MS sample fit. The clustering seems thus unaffected by the increased scatter in the Combined sample. We may then conclude that the MS fit gives a more robust model of the activity scaling, even for the full Combined sample, providing additional evidence for relating the knee to the axi- to non-axisymmetric transition.

\section{Discussion}

To fully understand the different activity levels in different types of stars, one would need to understand both how stellar dynamos depend on rotational properties and how their nonlinear saturation mechanism works. The latter is especially problematic for the following reasons. In mean-field dynamo models, very often, an ad hoc quenching formula is used, utilizing the assumption that the growing magnetic field starts influencing the flow field when the magnetic energy reaches equipartition with the kinetic energy of the flow \citep[see e.g.][and references therein]{Ch10}. Such an approach does not help in understanding how the saturation process occurs. There are also more physical attempts to use the magnetic helicity conservation law to derive a dynamic equation for the $\alpha$ effect \citep[see e.g.][and references therein]{BS05}. In this case the saturation level becomes dependent on various additional physical parameters, such as the magnetic Reynolds number, and the helicity fluxes out of the dynamo active domain. Unfortunately, these parameters are largely unknown, and in practice this approach only increases the number of unknowns in the problem.

Hence, the only remaining route are the so called direct numerical simulations where the full set of MHD equations is solved. This retains the Lorentz force feedback and allows for the magnetic helicity fluxes, that are thought to be vital in the nonlinear saturation process. The problem with this approach is that these models are still far removed from the realistic parameter regime of stars; most notably the viscosity, resistivity and thermal conduction being far increased from the real objects. There are some works that have already studied the rotation dependence of the dynamo solutions either in axisymmetric wedges \citep{KMCWB13,KKKOWB17,W18,WK20}, latitudinal wedges covering the full longitudinal extent \citep{VWK18,VK20}, or in full spheres \citep{NBBMT13,SBCBN17}. In such models the axi- to non-axisymmetric transition is seen \citep[e.g.][but requiring high enough resolution]{VWK18}, lending support to connecting the observed knee with this transition. The transition point is, however, located still at too low rotation rates in comparison to observations. Recently, \citet{VK20} have shown that an improved description of the heat conduction used in the model can push the transition into a more realistic direction.

The increase of the magnetic energy in the models as a function of rotation is also not correctly captured, unless the magnetic energy is normalized with the kinetic energy \citep[see e.g.][]{VWK18,W18,WK20}. In this case the normalized energy is seen increasing roughly proportional to the Coriolis number, in rough agreement with the observations. However, no such knee-point, as observationally confirmed, can be seen in these simulations. On the contrary, the increase of the magnetic to kinetic energy ratio occurs smoothly over the axi- to non-axisymmetric dynamo mode transition. These discrepancies could indicate that the models do not yet take correctly into account the rotational dependence of the critical Rayleigh number for the onset of convection: the more rapid the rotation, the harder convection becomes to excite. Unless the thermal conduction is decreased correspondingly when rotation rate is increased, which is usually not done in the modelling attempts, the energy in the convective motions might become
underestimated in the rapidly rotating cases.

To this day, the understanding of the rotation--activity scaling remains incomplete, as a whole. There is evidence that the RI regime itself has a shallow nonzero slope \citep{Reiners2014RotationActivity,Shulyak2019MDwarfFields,Magaudda2020RotationActivity}, which would make the transition between it and the RD regime qualitatively similar to the knee-point, discussed in this paper. No satisfactory physical explanation exists yet for the RD--RI transition. At the slow rotation end of the activity scaling, \cite{Brandenburg2018EnhancedActivity} reported potential activity enhancement, which they connected to a transition between solar and anti-solar differential rotation. This suggestion relies on results from global magnetoconvection studies, which indicate that the dynamo saturation level is enhanced in the anti-solar differential rotation regime w.r.t.~the solar-like regime \citep{KKKBOP15}. This effect is not unambiguously detected by all modelling efforts \citep[see e.g.][]{VWK18}. While this feature does have a proposed physical explanation, it relies so far on only a handful of stars and needs to be more securely verified observationally. Gaining a solid grasp of all of these features of the rotation--activity scaling is crucially connected to fully understanding the nonlinear saturation mechanism of stellar dynamos.

\section{Conclusions}

Our results provide strong evidence of the rotation--activity relation not being smooth in the rotation-dependent regime but rather having a localized break at mid-activity levels. For the main sequence stars, a two-piece power law model, with distinctly different slopes on either sides of this knee-point, clearly describes the activity data better than an often-used, smooth, exponential model. According to our model fit comparison, including giant stars in the sample would make the exponential model the preferred one. Our Gaussian clustering analysis, however, finds the knee-point regardless of whether the giant stars are considered or not. Since power law relations are also physically expected to arise from the MHD equations, unlike exponential ones, we conclude the two-piece power law model to be a more accurate description of the activity scaling relation.

We argue that the break in the activity scaling can be interpreted as a transition between two dynamo regimes, dominating at different rotation rates. A good candidate for identifying with this transition is the shift from axi- to non-axisymmetric magnetic configurations. This transition has been observed to occur at nearly the same activity levels and Rossby numbers in both spot activity \citep{Lehtinen2016Photometry} and surface magnetic fields \citep{See2016FieldTopology} as we find here for the knee-point.

\pagebreak

\acknowledgments

J.J.L.~acknowledges the support of the Independent Max Planck Research Group ``SOLSTAR''.
M.J.K.~acknowledges the support of the Academy of Finland ReSoLVE Centre of Excellence (grant number 307411).
N.O.~acknowledges the support of the SOLARNET H2020 project grant no.~824135.
F.S.~acknowledges the support of the German space agency (Deutsches Zentrum f\"ur Luft- und Raumfahrt) under PLATO Data Center grant 50OO1501.
This project has received funding from the European Research Council (ERC) under the European Union's Horizon 2020 research and innovation programme (Project UniSDyn, grant agreement n:o 818665).

The chromospheric activity data derive from the Mount Wilson Observatory HK Project, which was supported by both public and private funds through the Carnegie Observatories, the Mount Wilson Institute, and the Harvard-Smithsonian Center for Astrophysics starting in 1966 and continuing for over 36 years.  These data are the result of the dedicated work of O.~Wilson, A.~Vaughan, G.~Preston, D.~Duncan, S.~Baliunas, and many others.

\bibliography{knee}{}

\begin{thebibliography}{}
\expandafter\ifx\csname natexlab\endcsname\relax\def\natexlab#1{#1}\fi
\providecommand{\url}[1]{\href{#1}{#1}}
\providecommand{\dodoi}[1]{doi:~\href{http://doi.org/#1}{\nolinkurl{#1}}}
\providecommand{\doeprint}[1]{\href{http://ascl.net/#1}{\nolinkurl{http://ascl.net/#1}}}
\providecommand{\doarXiv}[1]{\href{https://arxiv.org/abs/#1}{\nolinkurl{https://arxiv.org/abs/#1}}}

\bibitem[{{Astudillo-Defru} {et~al.}(2017){Astudillo-Defru}, {Delfosse},
  {Bonfils}, {Forveille}, {Lovis}, \&
  {Rameau}}]{AstudilloDefru2017MDwarfActivity}
{Astudillo-Defru}, N., {Delfosse}, X., {Bonfils}, X., {et~al.} 2017, \aap, 600,
  A13

\bibitem[{{Auri{\`e}re} {et~al.}(2015){Auri{\`e}re}, {Konstantinova-Antova},
  {Charbonnel}, {Wade}, {Tsvetkova}, {Petit}, {Dintrans}, {Drake}, {Decressin},
  {Lagarde}, {Donati}, {Roudier}, {Ligni{\`e}res}, {Schr{\"o}der},
  {Landstreet}, {L{\`e}bre}, {Weiss}, \&
  {Zahn}}]{Auriere2015GiantMagneticFields}
{Auri{\`e}re}, M., {Konstantinova-Antova}, R., {Charbonnel}, C., {et~al.} 2015,
  \aap, 574, A90

\bibitem[{Barber(2012)}]{BarberBRML2012}
Barber, D. 2012, {Bayesian Reasoning and Machine Learning} ({Cambridge
  University Press})

\bibitem[{{Basri}(1987)}]{Basri1987BinaryActivity}
{Basri}, G. 1987, \apj, 316, 377

\bibitem[{{Brandenburg} \& {Giampapa}(2018)}]{Brandenburg2018EnhancedActivity}
{Brandenburg}, A., \& {Giampapa}, M.~S. 2018, \apjl, 855, L22

\bibitem[{{Brandenburg} \& {Subramanian}(2005)}]{BS05}
{Brandenburg}, A., \& {Subramanian}, K. 2005, \physrep, 417, 1

\bibitem[{{Charbonneau}(2010)}]{Ch10}
{Charbonneau}, P. 2010, Liv. Rev. Sol. Phys., 7, 3

\bibitem[{{Charbonnel} {et~al.}(2017){Charbonnel}, {Decressin}, {Lagarde},
  {Gallet}, {Palacios}, {Auri{\`e}re}, {Konstantinova-Antova}, {Mathis},
  {Anderson}, \& {Dintrans}}]{Charbonnel_ea:2017}
{Charbonnel}, C., {Decressin}, T., {Lagarde}, N., {et~al.} 2017, \aap, 605,
  A102

\bibitem[{{Douglas} {et~al.}(2014){Douglas}, {Ag{\"u}eros}, {Covey}, {Bowsher},
  {Bochanski}, {Cargile}, {Kraus}, {Law}, {Lemonias}, {Arce}, {Fierroz}, \&
  {Kundert}}]{Douglas2014ClusterActivity}
{Douglas}, S.~T., {Ag{\"u}eros}, M.~A., {Covey}, K.~R., {et~al.} 2014, \apj,
  795, 161

\bibitem[{{Folsom} {et~al.}(2018){Folsom}, {Bouvier}, {Petit}, {L{\`e}bre},
  {Amard}, {Palacios}, {Morin}, {Donati}, \&
  {Vidotto}}]{Folsom2018FieldEvolution}
{Folsom}, C.~P., {Bouvier}, J., {Petit}, P., {et~al.} 2018, \mnras, 474, 4956

\bibitem[{{Foreman-Mackey} {et~al.}(2013){Foreman-Mackey}, {Hogg}, {Lang}, \&
  {Goodman}}]{ForemanMackey2013emcee}
{Foreman-Mackey}, D., {Hogg}, D.~W., {Lang}, D., \& {Goodman}, J. 2013, \pasp,
  125, 306

\bibitem[{{Gilliland}(1985)}]{Gilliland1985RotationActivity}
{Gilliland}, R.~L. 1985, \apj, 299, 286

\bibitem[{{Goodman} \& {Weare}(2010)}]{Goodman2010EnsembleSamplers}
{Goodman}, J., \& {Weare}, J. 2010, Communications in Applied Mathematics and
  Computational Science, 5, 65

\bibitem[{{Hempelmann} {et~al.}(1995){Hempelmann}, {Schmitt}, {Schultz},
  {Ruediger}, \& {Stepien}}]{Hempelmann1995CoronalRotationActivity}
{Hempelmann}, A., {Schmitt}, J.~H.~M.~M., {Schultz}, M., {Ruediger}, G., \&
  {Stepien}, K. 1995, \aap, 294, 515

\bibitem[{{K{\"a}pyl{\"a}} {et~al.}(2017){K{\"a}pyl{\"a}}, {K{\"a}pyl{\"a}},
  {Olspert}, {Warnecke}, \& {Brandenburg}}]{KKKOWB17}
{K{\"a}pyl{\"a}}, P.~J., {K{\"a}pyl{\"a}}, M.~J., {Olspert}, N., {Warnecke},
  J., \& {Brandenburg}, A. 2017, \aap, 599, A4

\bibitem[{{K{\"a}pyl{\"a}} {et~al.}(2013){K{\"a}pyl{\"a}}, {Mantere}, {Cole},
  {Warnecke}, \& {Brandenburg}}]{KMCWB13}
{K{\"a}pyl{\"a}}, P.~J., {Mantere}, M.~J., {Cole}, E., {Warnecke}, J., \&
  {Brandenburg}, A. 2013, \apj, 778, 41

\bibitem[{{Karak} {et~al.}(2015){Karak}, {K{\"a}pyl{\"a}}, {K{\"a}pyl{\"a}},
  {Brandenburg}, {Olspert}, \& {Pelt}}]{KKKBOP15}
{Karak}, B.~B., {K{\"a}pyl{\"a}}, P.~J., {K{\"a}pyl{\"a}}, M.~J., {et~al.}
  2015, \aap, 576, A26

\bibitem[{{Kiraga} \& {St\c{e}pie\'n}(2007)}]{Kiraga2007RotationActivity}
{Kiraga}, M., \& {St\c{e}pie\'n}, K. 2007, \actaa, 57, 149

\bibitem[{{Kochukhov} {et~al.}(2020){Kochukhov}, {Hackman}, {Lehtinen}, \&
  {Wehrhahn}}]{Kochukhov2020HiddenFields}
{Kochukhov}, O., {Hackman}, T., {Lehtinen}, J.~J., \& {Wehrhahn}, A. 2020,
  \aap, 635, A142

\bibitem[{{Lehtinen} {et~al.}(2011){Lehtinen}, {Jetsu}, {Hackman}, {Kajatkari},
  \& {Henry}}]{Lehtinen2012CPS}
{Lehtinen}, J., {Jetsu}, L., {Hackman}, T., {Kajatkari}, P., \& {Henry}, G.~W.
  2011, \aap, 527, A136

\bibitem[{{Lehtinen} {et~al.}(2016){Lehtinen}, {Jetsu}, {Hackman}, {Kajatkari},
  \& {Henry}}]{Lehtinen2016Photometry}
---. 2016, \aap, 588, A38

\bibitem[{{Lehtinen} {et~al.}(2020{\natexlab{a}}){Lehtinen}, {Spada},
  {K{\"a}pyl{\"a}}, {Olspert}, \&
  {K{\"a}pyl{\"a}}}]{Lehtinen2020RotationActivity}
{Lehtinen}, J.~J., {Spada}, F., {K{\"a}pyl{\"a}}, M.~J., {Olspert}, N., \&
  {K{\"a}pyl{\"a}}, P.~J. 2020{\natexlab{a}}, Nature Astronomy, 4, 658

\bibitem[{{Lehtinen} {et~al.}(2020{\natexlab{b}}){Lehtinen}, {Spada},
  {K{\"a}pyl{\"a}}, {Olspert}, \& {K{\"a}pyl{\"a}}}]{Lehtinen2020Data}
---. 2020{\natexlab{b}}, VizieR Online Data Catalog (other), 0610,
  J/other/NatAs/4

\bibitem[{{Magaudda} {et~al.}(2020){Magaudda}, {Stelzer}, {Covey}, {Raetz},
  {Matt}, \& {Scholz}}]{Magaudda2020RotationActivity}
{Magaudda}, E., {Stelzer}, B., {Covey}, K.~R., {et~al.} 2020, \aap, 638, A20

\bibitem[{{Mamajek} \& {Hillenbrand}(2008)}]{Mamajek2008RotationActivity}
{Mamajek}, E.~E., \& {Hillenbrand}, L.~A. 2008, \apj, 687, 1264

\bibitem[{{Mittag} {et~al.}(2018){Mittag}, {Schmitt}, \&
  {Schr{\"o}der}}]{Mittag2018RotationActivity}
{Mittag}, M., {Schmitt}, J.~H.~M.~M., \& {Schr{\"o}der}, K.~P. 2018, \aap, 618,
  A48

\bibitem[{{Nelson} {et~al.}(2013){Nelson}, {Brown}, {Brun}, {Miesch}, \&
  {Toomre}}]{NBBMT13}
{Nelson}, N.~J., {Brown}, B.~P., {Brun}, A.~S., {Miesch}, M.~S., \& {Toomre},
  J. 2013, \apj, 762, 73

\bibitem[{{Newton} {et~al.}(2017){Newton}, {Irwin}, {Charbonneau}, {Berlind},
  {Calkins}, \& {Mink}}]{Newton2017MDwarfActivity}
{Newton}, E.~R., {Irwin}, J., {Charbonneau}, D., {et~al.} 2017, \apj, 834, 85

\bibitem[{{Noyes} {et~al.}(1984){Noyes}, {Hartmann}, {Baliunas}, {Duncan}, \&
  {Vaughan}}]{Noyes1984Activity}
{Noyes}, R.~W., {Hartmann}, L.~W., {Baliunas}, S.~L., {Duncan}, D.~K., \&
  {Vaughan}, A.~H. 1984, \apj, 279, 763

\bibitem[{{Olspert} {et~al.}(2018){Olspert}, {Lehtinen}, {K{\"a}pyl{\"a}},
  {Pelt}, \& {Grigorievskiy}}]{Olspert2018Cycles}
{Olspert}, N., {Lehtinen}, J.~J., {K{\"a}pyl{\"a}}, M.~J., {Pelt}, J., \&
  {Grigorievskiy}, A. 2018, \aap, 619, A6

\bibitem[{{Pizzolato} {et~al.}(2003){Pizzolato}, {Maggio}, {Micela},
  {Sciortino}, \& {Ventura}}]{Pizzolato2003CoronalRotationActivity}
{Pizzolato}, N., {Maggio}, A., {Micela}, G., {Sciortino}, S., \& {Ventura}, P.
  2003, \aap, 397, 147

\bibitem[{{Reiners} {et~al.}(2009){Reiners}, {Basri}, \&
  {Browning}}]{Reiners2009FluxSaturation}
{Reiners}, A., {Basri}, G., \& {Browning}, M. 2009, \apj, 692, 538

\bibitem[{{Reiners} {et~al.}(2014){Reiners}, {Sch{\"u}ssler}, \&
  {Passegger}}]{Reiners2014RotationActivity}
{Reiners}, A., {Sch{\"u}ssler}, M., \& {Passegger}, V.~M. 2014, \apj, 794, 144

\bibitem[{{Rutten}(1987)}]{Rutten1987RotationActivity}
{Rutten}, R.~G.~M. 1987, \aap, 177, 131

\bibitem[{{Saar}(2001)}]{Saar2001MagneticFields}
{Saar}, S.~H. 2001, in Astronomical Society of the Pacific Conference Series,
  Vol. 223, 11th Cambridge Workshop on Cool Stars, Stellar Systems and the Sun,
  ed. R.~J. {Garcia Lopez}, R.~{Rebolo}, \& M.~R. {Zapaterio Osorio}, 292

\bibitem[{{Schr{\"o}der} {et~al.}(2018){Schr{\"o}der}, {Schmitt}, {Mittag},
  {G{\'o}mez Trejo}, \& {Jack}}]{Schroder2018GiantActivity}
{Schr{\"o}der}, K.~P., {Schmitt}, J.~H.~M.~M., {Mittag}, M., {G{\'o}mez Trejo},
  V., \& {Jack}, D. 2018, \mnras, 480, 2137

\bibitem[{{See} {et~al.}(2016){See}, {Jardine}, {Vidotto}, {Donati}, {Boro
  Saikia}, {Bouvier}, {Fares}, {Folsom}, {Gregory}, {Hussain}, {Jeffers},
  {Marsden}, {Morin}, {Moutou}, {do Nascimento}, {Petit}, \&
  {Waite}}]{See2016FieldTopology}
{See}, V., {Jardine}, M., {Vidotto}, A.~A., {et~al.} 2016, \mnras, 462, 4442

\bibitem[{{Shulyak} {et~al.}(2019){Shulyak}, {Reiners}, {Nagel}, {Tal-Or},
  {Caballero}, {Zechmeister}, {B{\'e}jar}, {Cort{\'e}s-Contreras}, {Martin},
  {Kaminski}, {Ribas}, {Quirrenbach}, {Amado}, {Anglada-Escud{\'e}}, {Bauer},
  {Dreizler}, {Guenther}, {Henning}, {Jeffers}, {K{\"u}rster}, {Lafarga},
  {Montes}, {Morales}, \& {Pedraz}}]{Shulyak2019MDwarfFields}
{Shulyak}, D., {Reiners}, A., {Nagel}, E., {et~al.} 2019, \aap, 626, A86

\bibitem[{{Spada} {et~al.}(2017){Spada}, {Demarque}, {Kim}, {Boyajian}, \&
  {Brewer}}]{Spada_ea2017}
{Spada}, F., {Demarque}, P., {Kim}, Y.-C., {Boyajian}, T.~S., \& {Brewer},
  J.~M. 2017, \apj, 838, 161

\bibitem[{{Spada} {et~al.}(2013){Spada}, {Demarque}, {Kim}, \&
  {Sills}}]{Spada_ea2013}
{Spada}, F., {Demarque}, P., {Kim}, Y.~C., \& {Sills}, A. 2013, \apj, 776, 87

\bibitem[{{St\c{e}pie\'n}(1994)}]{Stepien1994RotationActivity}
{St\c{e}pie\'n}, K. 1994, \aap, 292, 191

\bibitem[{{Stoica} \& {Selen}(2004)}]{Stoica2004ModelSelection}
{Stoica}, P., \& {Selen}, Y. 2004, IEEE Signal Processing Magazine, 21, 36

\bibitem[{{Strugarek} {et~al.}(2017){Strugarek}, {Beaudoin}, {Charbonneau},
  {Brun}, \& {do Nascimento}}]{SBCBN17}
{Strugarek}, A., {Beaudoin}, P., {Charbonneau}, P., {Brun}, A.~S., \& {do
  Nascimento}, J.-D. 2017, Science, 357, 185

\bibitem[{{Su{\'a}rez Mascare{\~n}o} {et~al.}(2016){Su{\'a}rez Mascare{\~n}o},
  {Rebolo}, \& {Gonz{\'a}lez Hern{\'a}ndez}}]{SuarezMascareno2016CycleRotation}
{Su{\'a}rez Mascare{\~n}o}, A., {Rebolo}, R., \& {Gonz{\'a}lez Hern{\'a}ndez},
  J.~I. 2016, \aap, 595, A12

\bibitem[{{Vidotto} {et~al.}(2014){Vidotto}, {Gregory}, {Jardine}, {Donati},
  {Petit}, {Morin}, {Folsom}, {Bouvier}, {Cameron}, {Hussain}, {Marsden},
  {Waite}, {Fares}, {Jeffers}, \& {do
  Nascimento}}]{Vidotto2014RotationMagnetism}
{Vidotto}, A.~A., {Gregory}, S.~G., {Jardine}, M., {et~al.} 2014, \mnras, 441,
  2361

\bibitem[{{Vilhu}(1984)}]{Vilhu1984MagneticActivity}
{Vilhu}, O. 1984, in ESA Special Publication, Vol. 218, Fourth European IUE
  Conference, ed. E.~{Rolfe}, 239--242

\bibitem[{{Viviani} \& {K{\"a}pyl{\"a}}(2021)}]{VK20}
{Viviani}, M., \& {K{\"a}pyl{\"a}}, M.~J. 2021, \aap, 645, A141

\bibitem[{{Viviani} {et~al.}(2018){Viviani}, {Warnecke}, {K{\"a}pyl{\"a}},
  {K{\"a}pyl{\"a}}, {Olspert}, {Cole-Kodikara}, {Lehtinen}, \&
  {Brandenburg}}]{VWK18}
{Viviani}, M., {Warnecke}, J., {K{\"a}pyl{\"a}}, M.~J., {et~al.} 2018, \aap,
  616, A160

\bibitem[{{Warnecke}(2018)}]{W18}
{Warnecke}, J. 2018, \aap, 616, A72

\bibitem[{{Warnecke} \& {K{\"a}pyl{\"a}}(2020)}]{WK20}
{Warnecke}, J., \& {K{\"a}pyl{\"a}}, M.~J. 2020, \aap, 642, A66

\bibitem[{{Wilson}(1978)}]{Wilson1978MWOHK}
{Wilson}, O.~C. 1978, \apj, 226, 379

\bibitem[{{Wright} {et~al.}(2018){Wright}, {Newton}, {Williams}, {Drake}, \&
  {Yadav}}]{Wright2018MDwarfActivity}
{Wright}, N.~J., {Newton}, E.~R., {Williams}, P. K.~G., {Drake}, J.~J., \&
  {Yadav}, R.~K. 2018, \mnras, 479, 2351

\end{thebibliography}
\bibliographystyle{aasjournal}

\appendix

\begin{deluxetable*}{lccccc|cc}
\tablecaption{Fit coefficients of the power law model when including the $\log R'_{\rm HK}$ uncertainties\label{tab:fiterr}}
\tablehead{& $\log C_1$ & $\beta_1$ & $\beta_2$ & $\log {\rm Ro}_0$ & $\sigma$ & $\log R'_{\rm HK,0}$ & ${\rm Ro}_0$}
\startdata
MS & $-4.745^{+0.054}_{-0.048}$ & $-0.458^{+0.063}_{-0.060}$ & $-1.227^{+0.148}_{-0.188}$ & $-0.453^{+0.055}_{-0.072}$ & $0.084^{+0.009}_{-0.008}$ & $-4.537^{+0.054}_{-0.048}$ & $0.353^{+0.048}_{-0.054}$ \\
Combined & $-4.807^{+0.038}_{-0.036}$ & $-0.516^{+0.061}_{-0.057}$ & $-1.502^{+0.278}_{-0.299}$ & $-0.311^{+0.030}_{-0.053}$ & $0.119^{+0.010}_{-0.009}$ & $-4.646^{+0.040}_{-0.032}$ & $0.489^{+0.035}_{-0.056}$ \\
\enddata
\end{deluxetable*}

\section{Data tables}
\label{sect:data}

The data of the rotation--activity relations analysed in this study can be retrieved from the electronic tables published by \citet{Lehtinen2020Data}. These tables contain estimated values and uncertainties for all the derived parameters used for constructing the rotation--activity relations ($P_{\rm rot}$, $\tau_{\rm c}$, $\rm Ro$, $\log R'_{\rm HK}$, and the MS/giant evolutionary status), as well as the stellar astrophysical parameters used in the derivation of the rotation--activity parameters, including references for their original sources. For $P_{\rm rot}$ and $\log R'_{\rm HK}$ the reported error estimates represent the formal internal errors of the long term average period and chromospheric activity values. For $\tau_{\rm c}$ the uncertainty estimates were derived from the range of values obtained from stellar models with three different metallicities. The uncertainties of $\rm Ro$ were then derived from the uncertainties of both $P_{\rm rot}$ and $\tau_{\rm c}$ by propagation of uncertainty.

\section{Regression fits with included $\log R'_{\rm HK}$ uncertainties}
\label{sect:errfit}

In order to check the effect of the internal data errors to our model fit, we performed a modified regression of the two-piece power law model (Eq.~\ref{eq:model}), by including the individual observational errors of the average stellar $\log R'_{\rm HK}$ values. To achieve this, we used a modified scatter parameter $\sigma_i'^2 = \sigma^2 + \epsilon_i^2$ for the likelihood function Eq.~\ref{eq:likelihood}, which consists of the overall scatter parameter $\sigma$ and the individual uncertainty $\epsilon_i$ of each $\log R'_{\rm HK}$ measurement. Apart from this, we performed the regression in exactly the same way as the two-piece power law model regression described in Sect.~\ref{sect:fit}.

The parameter estimates resulting from the modified regression are listed in Table \ref{tab:fiterr}. The modified regression produces thus essentially identical results to our main regression. This is expected, since for each star the long term $\log R'_{\rm HK}$ averages are known to a far higher precision than the external scatter of the stellar activities, i.e. $\epsilon_i \ll \sigma$ for all stars. The same is also implied for the vanishing effect of the $\rm Ro$ uncertainties, since for all the main sequence and giant stars included in our analysis the estimated $\rm Ro$ uncertainties are much smaller than the external scatter along the $\rm Ro$ axis \citep[see][]{Lehtinen2020RotationActivity}. This validates our choice for the simpler setup for the regressions presented in Sect.~\ref{sect:fit}, which only use a single uniform scatter parameter $\sigma$ instead of considering the individual data errors.

\end{document}